# The first droplet in a cloud chamber track

Jonathan F. Schonfeld
Center for Astrophysics | Harvard and Smithsonian
Cambridge, Massachusetts, USA
jschonfeld@cfa.harvard.edu
ORCID ID# 0000-0002-8909-2401

**Abstract** In a cloud chamber, the quantum measurement problem amounts to explaining the first droplet in a charged-particle track; subsequent droplets are explained by Mott's 1929 wave-theoretic argument about collision-induced wavefunction collimation. I formulate a mechanism for how the first droplet in a cloud chamber track arises, making no reference to quantum measurement axioms. I look specifically at tracks of charged particles emitted in the simplest slow decays, because I can reason about rather than guess the form that wave packets take. The first visible droplet occurs when a randomly occurring, barely-subcritical vapor droplet is pushed past criticality by ionization triggered by the faint wavefunction of the emitted charged particle. This is possible because potential energy incurred when an ionized vapor molecule polarizes the other molecules in a droplet can balance the excitation energy needed for the emitted charged particle to create the ion in the first place. This degeneracy is a singular condition for Coulombic scattering, leading to infinite or near-infinite ionization cross sections, and from there to an emergent Born rule in position space, but not an operator projection as in the projection postulate. Analogous mechanisms may explain canonical quantum measurement behavior in detectors such as ionization chambers, proportional counters, photomultiplier tubes or bubble chambers. This work is important because attempts to understand canonical quantum measurement behavior and its limitations have become urgent in view of worldwide investment in quantum computing and in searches for super-rare processes (e.g., proton decay).

**Acknowledgements** I acknowledge productive interactions with Jacob Barandes, John Clauser, Steve Libby, Will Oliver, Hossein Sadeghpour, John Stack, the journal's reviewers, and the members of the Foundations Seminar of the Black Hole Initiative at Harvard.

**Declarations** The author did not receive support from any organization for the submitted work. The author has no relevant financial or non-financial interests to disclose

## 1. Introduction

The world is investing heavily in quantum computing [1], and the entire theoretical edifice of quantum computing rests squarely on the shoulders of the quantum measurement axioms. Specifically, quantum computing algorithms make heavy use of the projection postulate [2], the axiom that every measurement is strictly equivalent to random application of one of a set of mathematical projection operators, with probability governed by the Born rule.

We talk about a measurement *problem* [3] for several reasons. First, because the measurement axioms seem to exist in isolation from all other acknowledged laws of nature. As far as we know, the axioms play no role whatsoever in the microscopic dynamics of anything, let alone measurement devices. One invokes them only to leapfrog what should be deep analyses of measurement device internals, even though there is a virtual infinity of measurement device conceptual designs. Second, because no one has succeeded in deriving the behavior specified by the measurement axioms from accepted theories (or even heuristic models) of the microscopic dynamics of measurement devices. Third, because the projection postulate – formalizing the Copenhagen interpretation – involves manifestly non-unitary transformations, in jarring contrast to the unitary Schroedinger equation that the postulate itself assumes governs all behavior between measurements.

Since the founding of quantum mechanics, a great many physicists have proceeded as if the projection postulate or other measurement axioms are true under all circumstances and to infinitely many decimal places. But this postulate, and the Copenhagen interpretation before it, were formulated at a time when experimental and theoretical tools to probe them at what we now consider exquisite precision, and at exquisite spatial and temporal resolution, didn't exist. From our modern vantage point, it seems fair to wonder if this early formalization was premature. If so, then it's also fair to ask if the very basis of quantum computing breaks down, and how close the point of breakdown is to the present state of quantum computing technology. In light of current worldwide investment, these questions are urgent, not merely academic.

The projection postulate – or analogous formulations in other quantum interpretations (surveyed in [3]) – has enjoyed tremendous staying power over many decades because no violation has been recorded; but also because of its conceptual simplicity and the way it lends itself to elegant mathematical formalism and theorem proving. But these are not reasons to treat the postulate as dogma, especially when physicists otherwise make their livings questioning assumptions about all sorts of things, both fundamental and practical.

So, *is* the projection postulate or any related measurement axiom fundamentally and literally true if you look closely enough? In this paper, I will attempt to analyze the internal dynamics of a specific real single-quantum detector, the cloud chamber. I will account for the outcome of a cloud-chamber measurement – without invoking measurement axioms – by combining simple Schroedinger equation argumentation with heuristic models of the bulk medium that would be non-controversial under any other circumstances. This construction will point to "no" in response to this paragraph's lead question. Specifically, it will support the conclusions that the actual physical process at the heart of a measurement is much more complex and situation-specific than a simple projection, and that the Born rule is both limited and emergent.

There is a modest literature that attempts to analyze measurement models to discover how canonical quantum measurement phenomenology might arise without resorting to measurement axioms. Several years ago, the authors of Ref. [3] looked in great detail at magnetic measurement of a single spin, but dealt almost exclusively with density matrices rather than single measurement trials. The extensive decoherence program [4] also deals with density matrices rather than single measurement trials, and as such is able to explain away interference terms, but is unable to describe the occurrence of individual outcomes. There have been attempts [5] to construe single quantum measurement trials as examples of spontaneous symmetry breaking, but their scenarios have so far been severely limited and the conclusions drawn by them are so far not backed up by a dynamical analysis. I myself formulated a resonance mechanism to explain the position-space Born rule in an idealized model of a particle detector composed of two-level atoms [6]. Reference [7] attempted a density-matrix analysis of an idealized one-dimensional cloud chamber.

The most celebrated such paper is Mott's 1929 analysis of cloud chamber tracks [8]. Mott attempted to explain the track-like nature of cloud chamber detections entirely via the Schroedinger equation. In particular, he showed that the scattered wave emerging from the first collision of a charged particle in the chamber must be strongly collimated, requiring all subsequent collisions (including ionizations) to lie in the narrow cone of collimation. As a result, for all practical purposes, the subsequent collisions occur as if the projectile is a point particle, and the naive semiclassical cross-section picture dictates collision statistics. But Mott treated the first scattering as a perturbation, leaving unexplained why the first collision doesn't leave behind a residual unscattered wave that's still spatially extended and can go on to start distinct tracks elsewhere. He didn't explain how a track starts in the first place, that is, how the first droplet comes about. So we are still left with no explanation for why, say, the s-wave alpha particle emitted by a single $U^{238}$ is observed as a single track pointing in a single direction, starting at a single time, rather than as a gradually revealed spherical haze.

In this paper, I formulate a mechanism for how the first droplet in a single charged-particle track arises, and in the process show how it follows an emergent position-space Born rule, without invoking measurement axioms. I look specifically at tracks of particles emitted in very simple slow decays (initial heavy particle -> final heavy particle + single light charged particle), because in that case I can reason about the forms of wave packets rather than rely on unsupported guesswork. "Slow" in this case means that the decay lifetime is much longer than the time it takes for an emitted particle to cross the detector. I distinguish between the Born rule in general and the position-space Born rule, because the former applies only when measurements are literally the same as projections, and that is not the case here.

I will shortly provide a thumbnail summary of the basic physics worked out in this paper. But first, as a guide to readers in the quantum-fundamentals community, I explain how this work relates to or contrasts with the expectations of the projection postulate or the Copenhagen interpretation:

- *Interpretation generally*: My aim in this paper is to identify physical mechanisms responsible for the apparent phenomenology represented by measurement axioms, and if possible to discover their situation-specific limitations. In order to do this I must assume a concrete phenomenological model of a measuring device, and some may object that this implicitly entails an underlying quantum interpretation of some kind. If one conflates "measurement axioms" with "quantum interpretation," then in this sense it could be cheating to derive measurement behavior if it's already implicit in the assumed device model. But I am claiming in this paper that canonical measurement behavior and its limitations can be derived from physical considerations that themselves make no *explicit* reference to measurements. If a quantum interpretation of some kind is necessary for asserting that, for example, droplets form spontaneously in supersaturated vapors whether we observe them or not, then, narrowly viewed, the most foundational result in this paper is that the interpretation must be framed *without* direct reference to measurement, in direct contrast to Copenhagen and Many Worlds, for example. I don't understand why interpretation-free unitary wavefunction evolution plain and simple *can't* account for all aspects of the world around us, but strictly speaking this paper may not actually require that.
- *Detector model:* The detector model in this paper is neither entirely classical nor entirely quantum mechanical. It assumes a classical picture of subcritical droplet formation occurring in the background because of thermal fluctuation, but it assumes a quantum-mechanical, discrete-state model of the molecule ionized by a passing charged particle.
- *Randomness:* Subcritical droplet formation in the background is assumed to occur randomly in space and time, presumably as a result of thermal fluctuations in a many-molecule medium. The probabilistic nature of cloud-chamber track formation is then not intrinsic to quantum mechanics but is a direct consequence of this thermal randomness, which I assume exists independent of any measurement axioms, i.e. independent of whether or not anyone observes it. At a fundamental level, the cloud chamber system is deterministic. Accordingly, all probabilities enunciated in this paper are meant to be frequentist.
- *Wavefunction collapse:* When detection occurs (when the first droplet is formed in a cloud chamber track), the quantum mechanical state of the overall system makes a rapid transition that is neither instantaneous nor a projection/collapse. The state's *emitted-particle* content evolves from a configuration symmetric about the initial heavy particle to one collimated along a specific radius pointing away from the heavy particle. The state's *detector-medium* content evolves from a particular configuration of subcritical droplets to a configuration of supercritical droplets along the emitted particle collimation axis. This overall transition can be one-to-one, consistent with unitarity, rather than many-to-one as required by projection, because there are plenty of particular degrees of freedom in the teeming cloud chamber medium. *If* we ignored the detector part of the quantum state, this would indeed look like random collapse and projection, but (see below) that's just not how the physics actually works.

- *Logical completeness:* In Section 4, in order to arrive at the position-space Born rule, I will find it necessary to invoke what may seem like a very weak version of the Born rule itself. This is the assumption that very small contributions to quantum amplitudes have negligible impact on experimental observations. Aside from risking circularity, this also suffers from my inability to quantify "very small." This is not a new problem. For example, Ref. [9] surveys a number of attempts to pair this assumption with the axiom that all measurements are projections. The present paper is different in making no recourse to projections. Foreshadowing the discussion at the end of Section 4, this small-amplitude assumption could very well be a worthy subject of theoretical research in its own right.

As indicated earlier, my demonstration combines simple Schroedinger equation argumentation with heuristic models of the bulk medium that would be non-controversial under any other circumstances. I cannot claim to have solved a particular case of the measurement problem in any rigorous sense. However, I can hope that the train of thought described here provides productive guideposts for how to prosecute such a rigorous solution.

Here is the basic physical picture that this paper will flesh out:

- *Decay scenario:* For the purposes of this paper, a decay system consists of an initial heavy, stationary particle; a final heavy, nearly stationary particle; and an emitted light charged particle. We shall refer to the initial heavy particle as "initial source," the final heavy particle as "final source," and the light charged particle as "emitted particle." The Hamiltonian spectrum of such a system consists of two categories of states. One category ("free") consists of many states, each representing a free emitted particle and the final source, but with a very small admixture of the undecayed initial source. The second category ("persistent") consists of a single state representing the initial source with an additive admixture that combines final source with an emitted particle with a localized spatial wavefunction that takes a universal form close to the decay source. We will see that, over long times, this single persistent state accounts for detector physics near the decay source, in particular inside the detector. It is the emitted-particle wavefunction of this persistent state that is left behind to interact with the cloud chamber medium after the spatial wavefront of the emitted particle speeds outward. But, in the persistent state, the emitted-particle wavefunction and the initial source are rigidly coupled by superposition: When one is consumed (see below) by an interaction within the detector, the other must be consumed with it. This is why an initial decay source is observed to disappear abruptly when a track forms (i.e., why one source can't make two tracks at two separate times).
- *Detector model:* A cloud chamber is an enclosure containing air supersaturated with a condensable vapor, which can be water but is more typically ethyl alcohol [10]. Conventional wisdom has it that when a charged particle passes through the chamber, it ionizes air molecules, and the ions nucleate visible vapor droplets. However, this assumes the charged particle wavefunction is very collimated (so the particle can be treated as a point), while the persistent wavefunction of an emitted charged particle near the initial source is not collimated in any meaningful sense.

- *First-droplet formation:* Instead, the persistent wavefunction interacts with a vapor droplet (occurring randomly due to thermal fluctuations) that is just barely sub-critical. A barely sub-critical droplet has a very large amplitude of interaction (see below) with the persistent wavefunction. As a result, even a very weak persistent wavefunction can provoke the subcritical droplet to grow quickly in a supercritical fashion and become visible. This interaction "consumes" the persistent-state wavefunction (wavefunction collapse) in the sense that this wavefunction (i.e. localized emitted particle and large admixture of initial state) makes an effectively complete transition to a free wavefunction (i.e. escaping emitted particle and small admixture of initial state). This transition is effectively complete because, in the presence of a droplet that's already formed, single-molecule ionization can proceed with very small energy loss, since ion-induced potential energy due to droplet polarization can nearly balance electron excitation energy. This near-degenerate situation can drive quantum Coulomb scattering cross sections to singularity [11]. Mott's arguments guarantee that a strongly collimated free wavefunction emerges from this transition. This is the microscopic origin of what one observes as the position-space Born rule.

For obvious reasons, it is highly desirable to extend these ideas to other types of measurement. I think this can be done for detectors that rely on gas-filled enclosures and electrified conductors, such as ionization chambers, proportional counters and photomultiplier tubes. My motivation for this conjecture is that, in algebraic form and order of magnitude, the ion-induced droplet potential in a cloud chamber is extremely similar to the ion-induced image-charge potential next to a conducting surface. It is vital to understand the limits of canonical quantum measurement phenomenology in this class of detector because it is so widespread. It is especially vital because such detectors are critical to searches for exceedingly rare processes such as proton decay. We need to know if we can no longer rely on canonical quantum measurement expectations when decays get too rare. If that's the case, we could be over- or under-estimating experimental bounds on the lifetimes of exceedingly rare decays. The ideas in this paper also hint at an approach to understanding the first bubble in a bubble chamber track.

Before proceeding to detailed argumentation, I offer two further comments to help place this work in a broader context. First, the constructions in this paper specifically address the case of a single emitted particle, so the situation covered by Bell's theorem would seem to be out of scope. However, it's easy to see how this work respects Bell's theorem. Imagine replacing "final source" by "vacuum," and "emitted particle" by "two entangled emitted particles exiting in opposite directions." Then if one particle has a collision that collimates its wavefunction, that collision must also collimate the entangled particle's wavefunction. This is precisely the "spooky action at a distance" that gives rise to Bell's theorem. Second, given the undifferentiated generality of the measurement axioms, one may have expected that explaining any part of them should not refer to interaction specifics. This is not the case here: the Coulomb nature of charged particle interactions is critical. We will return to this point at the end of Section 4. If the particle to be detected is neutral, then the arguments in this paper must refer to Coulomb interactions of whatever charged particles result from the neutral's collision with a detector molecule (for example, the helium and lithium ions emitted when $B^{10}$ absorbs a neutron in a $BF_3$ proportional counter).

The remainder of this paper is organized as follows. In Section 2, I specify the detailed cloud chamber phenomenology that needs explaining. In Section 3, I analyze emitted-particle wavefunctions. In section 4, I analyze the physics of subcritical vapor droplets and quantum Coulomb scattering from degenerate states as they relate to cloud chamber track formation. In section 5, I consider the case of detectors that rely on gas-filled chambers and electrified conductors, such as ionization chambers, proportional counters and photomultiplier tubes. In section 6, I briefly consider bubble chambers. I summarize results and discuss their significance in Section 7.

## 2. Cloud chamber detection phenomenology

It seems to me that any measurement-axiom-free line of argumentation must account for these basic facts about cloud chamber detection of a charged particle emitted by a decay:

1. *Finality:* An initial source decays only once: After a track appears, there's no more initial source left to stimulate further tracks.
2. *Suddenness:* An emitted-particle track appears abruptly, even when the decay lifetime itself is extremely long (4B years for $U^{238}$ -> $\alpha$ + $Th^{234}$).
3. *Randomness:* The origin of an emitted particle track is randomly distributed in space and time.
4. *Probability distribution:* The probability per volume and per time that an emitted particle track starts at a particular location and particular time is proportional to the absolute-value-squared of the emitted particle's spatial wavefunction in the persistent state at that location and time, and is otherwise independent of the details of the underlying decay process. But note: As far as I can tell, no experiment has actually demonstrated this, except possibly with respect to dependence on solid angle. Thus, I take item #4 as a good heuristic, but there is a real gap in the science here.

## 3. Hamiltonian structure of quantum decay; decay finality

In the popular imagination, we think of decay as something that happens abruptly to a metastable state: the metastable state persists for some time and then it's gone. It's a very different picture at the quantum level.

To start, let us consider a well-known model that can be solved exactly. In [12], Stey and Gibberd solved analytically for the unitary operator U that describes evolution of a point-like two-level atom coupled to the radiation field of a unidirectional photon in one-dimensional space. The photon serves the role of "emitted particle" in the language of Section 1. In the notation of [12], the important matrix elements are

$$U_{\varphi\varphi} = e^{-(v^2 - i\varepsilon_\varphi)t} \qquad (3.1)$$

$$U_{n\varphi} = -\frac{iv}{\sqrt{l}}\left[i(\varepsilon_n - \varepsilon_\varphi) - v^2\right]^{-1}\left\{e^{-i\varepsilon_n t} - e^{-(v^2 - i\varepsilon_\varphi)t}\right\}, \qquad (3.2)$$

where $|\varphi\rangle$ is the state of an excited atom (initial source, excitation energy $\varepsilon_\varphi$) and no photon (i.e. cross ($\otimes$) the photon vacuum). $|n\rangle$ is the state of ground-state atom (final source) and (cross ($\otimes$)) one photon of wavenumber $\pi n/l$. $2l$ is the length of the quantization box; $\varepsilon_n$ is $\pi n/l$ (energy can be negative in this model); $v$ is a real coupling constant; and the speed of light and the reduced Planck's constant are both normalized to unity. Recurrences (i.e. reflections or wraparounds from the edges of the quantization volume) are neglected for large $l$. (Clearly, in this case the decay e-folding rate is $\gamma = 2v^2$.) If a state is $|\varphi\rangle$ at time zero, then at time $t$ it's

$$|\varphi\rangle U_{\varphi\varphi} + \sum_n |n\rangle U_{n\varphi}. \qquad (3.3)$$

By Fourier transformation, the photon-mode sum becomes the spatial wavefunction

$$\psi(x) \equiv \sum_n (2l)^{-\frac{1}{2}} e^{-\frac{ix\pi n}{l}} U_{n\varphi} = -iv\sqrt{2}e^{(x-t)(v^2 - i\varepsilon_\varphi)}, \; 0 < x < t \qquad (3.4)$$
$$= 0 \text{ otherwise}$$

(after using contour integration for the large-quantization-volume (large-$l$) limit, and assuming $\mathrm{Im}\,\varepsilon_\varphi > 0$). Far from the decay source, this is a Lorentzian wavepacket, i.e. a wave front moving to the right at light speed, and tailing off to the left like $\exp[x(v^2 - i\varepsilon_n)]$. But near the origin, for slow decays (small $v^2$) it is basically the plane wave

$$-iv\sqrt{2}e^{-ix\varepsilon_\varphi}, \qquad (3.5)$$

scaled by the right-hand-side of Equation (3.1). In consequence, we can say that the decaying initial source crossed ($\otimes$) with photon vacuum is, at all times, joined by superposition to a final source crossed ($\otimes$) with an emitted photon that has a very particular spatial wavefunction. The initial-source ($\otimes$ photon vacuum) amplitude and the emitted-photon ($\otimes$ final source) spatial wavefunction close to the source evolve in lockstep – both scale with time as $\exp[-t(v^2 - i\varepsilon_\varphi)]$. Significantly, this lockstep configuration is precisely the persistent state introduced in Section 1, as we now explain.

The exact characteristic equation of the full Hamiltonian for the combined atom-photon system is [8]

$$E - \varepsilon_\varphi = v^2 \cot(lE). \qquad (3.6)$$

The corresponding eigenvector, up to normalization but otherwise without approximation, is [5]

$$|\varphi> + v\sqrt{l} \sum_n \frac{1}{El - n\pi} |n>. \tag{3.7}$$

For small $v$, an eigenvalue may be close to $n\pi/l$ with remainder $-v^2/l\varepsilon_\phi$, so the sum in Equation (3.7) is $O(1/v)$ and overwhelms the $|\phi>$ term. These are the free states introduced in Section 1. Or an eigenvalue may be close to $\varepsilon_\phi$, with remainder $v^2\cot(l\varepsilon_\phi)$, so the sum in Equation (3.7) is $O(v)$ and is accompanied by a non-negligible $|\phi>$ term (we ignore accidental resonance between $\varepsilon_\phi$ and some $n\pi/l$). In the latter case, the sum in Equation (3.7), Fourier transformed into position space, is

$$\psi(x) = \frac{v}{\sqrt{2}} \sum_n e^{-\frac{ix\pi n}{l}} \frac{1}{\varepsilon_\varphi l - n\pi} = \frac{v}{\sqrt{2}} [\cot(\varepsilon_\varphi l) - i] e^{-ix\varepsilon_\varphi} = -iv\sqrt{2} e^{-ix\varepsilon_\varphi}, \tag{3.8}$$

for $x>0$, and zero otherwise (the far right-hand-side is the infinite-$l$ limit for Im$\varepsilon_\phi$>0). This is the persistent state introduced in Section 1, and is exactly the same as Equation (3.5).

I will explain the significant of this shortly. First, let us extend this framework to three dimensions and a nonrelativistic emitted particle. The extension is intuitively obvious, even though I don't have an analytical calculation like Stey and Gibberd's to anchor it. The radial parts of amplitudes still evolve according to Equation (3.4), appropriately generalized from photon to massive nonrelativistic particle. Far from the source, the emitted particle is described by a spatial wavefront receding at $r \sim st$ (mass $m$, momentum $p=ms$), with thickness $2/\gamma$ in time and therefore $2s/\gamma$ in space. Close to the source, but beyond its tiny interaction radius, the emitted-particle spatial wavefunction should be (up to a time-dependent but irrelevant phase)

$$\psi(x) \sim -iY(\Omega) \sqrt{\frac{\gamma}{s}} \frac{1}{r} \exp\left\{\left(\frac{r}{s} - t\right)\left(\frac{\gamma}{2} - i\frac{ps}{\hbar}\right)\right\}, \tag{3.9}$$

where Y is a spherical harmonic and $\Omega$ is solid angle, and we have restored Planck's constant. [Implicitly, we have also assumed that kinetic energy $k^2/2m$ can be linearized around momentum $k=p$, just as in the usual textbook analysis of group vs. phase velocity.] In Gamow's heuristic theory of nuclear alpha decay [13], Equation (3.9) is the part of the steady-state alpha wavefunction outside the nucleus. The decaying initial source $|\phi>$ is the part of Gamow's wavefunction inside the nucleus.

All this is significant for the following reason. Conventionally, one parameterizes the interaction of charged particles with detector molecules via transition amplitudes between eigenstates of a fictitious "unperturbed" charged-particle Hamiltonian that ignores the detector. With no decay source, these eigenstates are the plane waves familiar from textbooks. With a decay source, these eigenstates are the free and persistent states introduced above. We have just seen that the decaying system inside a finite-size detector is well approximated by the persistent state for long times. Therefore, the quantum transition corresponding to detection must be from persistent state to a superposition of free states. The persistent state has a significant admixture of initial source; the initial-source admixture in the free-state superposition is negligible. This is why, *if* the persistent state makes this transition with negligible remainder (i.e. if it is "consumed"), a decay source can only be observed to decay once ("finality" in Section 2). This doesn't explain yet why such a transition is consuming and abrupt; that will the subject of the next section.

It will be useful in what follows to have noted here that, numerically, amplitudes for persistent-to-free transitions in a detector are determined only by the emitted-particle-wavefunction content of these states. This is because the decay source itself doesn't interact directly with the detector molecules in any significant way. For this reason, in what follows I will estimate persistent-to-free transition amplitudes using conventional scattering theory, without direct reference to the decay source. Rather than further explaining this point here, I refer the reader to the next section for clarification by example.

Equation (3.9) sets a scale for thinking about cloud chamber behavior. As an extreme but familiar physical example, consider $U^{238}$ alpha decay, for which $\gamma^{-1} \sim 4B$ years and $s \sim c/20$. Then, a mere centimeter from the decaying nucleus, the density $|\psi|^2$ is $O(10^{-20} \text{ m}^{-3})$. This persistent wavefunction, which I claim is the trigger for abrupt and incontrovertible track formation, seems very far from collimated.

## 4. Subcritical vapor droplets and degenerate Coulomb scattering

This lack of collimation leads me to question the plausibility that an emitted-particle wavefunction close to the initial source can prompt the first visible track droplet into existence "from scratch" by ionizing an isolated gas molecule. Furthermore, the abruptness of the appearance of a droplet implies an underlying process with a very large rate, and there are simply no very large rates in sight without some special amplification mechanism. I therefore make the working hypothesis that the role of the wavefunction is to provide a mere "final nudge" to a just-barely-subcritical vapor droplet that's already formed by garden-variety random thermal fluctuations, in the hope that rates for underlying ionization processes at the molecular level are enhanced dramatically for near-critical droplets.

Basics of droplet formation

To play out this working hypothesis, let us begin by reviewing the conventional picture of how subcritical droplets form and how their growth is accelerated by ion formation (classical nucleation theory). (There is a considerable literature [14] criticizing, deconstructing and revising the conventional picture, but I think the conventional view will suffice for our needs.) In a supersaturated gas, vapor molecules cluster randomly (anticipating "randomness" in Section 2) into droplets. There is a critical size below which a droplet evaporates, and above which it grows rapidly to being visible. Subcritical droplet statistics are determined by a free energy $\Delta G$, which is a function of droplet population $n$ (not to be confused with the photon index in Section 3), and the concentration of droplets of population $n$ is proportional to the Boltzmann weight exp(– $\Delta G_n/k_BT$). In the absence of external influence, including ionization, free energy $\Delta G_n$ has a maximum, reflecting a competition between clustering and evaporation, which balance when a droplet is critical. For future use, let $\tau$ be the typical evaporation time, presumed short, for such a droplet not far below critical, and let $\rho dR$ be the number of such droplets with radius between $R$ and $R+dR$, per unit volume and unit time. As indicated earlier, I assume that clustering and thermal fluctuation are characteristic of the cloud chamber medium whether anyone observes it or not, i.e. their occurrence follows from conventional continuous-time dynamics; and that intrinsically discontinuous processes such as axiomatic measurement projections play no role whatsoever.

In what follows we work in terms of $R$ rather than $n$, because $R$ varies continuously while $n$ does not. A droplet of molecules with a fixed $n$ can take a continuous range of $R$ values because of conformational variability.

If there is an ion of charge $Q$ at the center of the droplet, the free energy is supplemented by the term (Kelvin-Thomson model [15])

$$g(R) \equiv \frac{Q^2}{2}(1 - 1/\varepsilon)\left(\frac{1}{R} - \frac{1}{R_i}\right), \tag{4.1}$$

where $\varepsilon$ is droplet dielectric constant (generally $\gg 1$), $R$ is the radius of the droplet, $R_i$ is the effective radius of the ion, and I assume charge is normalized as is customary in particle physics, where the Coulomb potential takes the form $Q_1Q_2/r$. The quantity $g(R)$ is clearly negative. It reduces the free energy maximum so that it becomes much easier for thermal fluctuations to surmount the droplet-formation barrier. That's how ionization nucleates supercritical droplet formation.

Equation (4.1) may seem strange because we are used to thinking that the obvious ionization targets in a cloud chamber are $O_2$ or $N_2$ molecules, which dominate the molecular population. If that were the case, the ion that Equation (4.1) refers to would most likely be situated just *outside* the droplet rather than at its center, with a completely different functional form for $g(R)$, and with a smaller numerical value, given $Q$ and $R_i$. But where there's a droplet, the predominating target for ionization is what makes up the droplet itself, and the dominant specific molecule would be at the location that extremizes $g$, i.e. at the droplet's center.

If the molecule to be ionized starts out with no accompanying droplet ($R = R_i$), then in our decay situation it's very hard in practice to trigger droplet formation since the bare ionization amplitude of our emitted-particle projectile is small and its persistent wavefunction is very weak. The situation changes if the molecule to be ionized is located in a non-null droplet ($R > R_i$), because that changes the energetics, which can have a profound impact on interaction amplitudes. Let $\Delta E$ be the emitted-particle energy change due to creating an ion (our sign convention is that energy loss is $\Delta E < 0$); $E_I$ ($>0$) be ionization potential; and $E_e$ be the kinetic energy of the ejected electron (we assume single-electron ionization). Then we have

$$0 = \Delta E + E_I + E_e + g(R) \equiv \Delta E + E_e + \delta. \tag{4.2}$$

As a droplet grows from nothing, $\delta$ starts positive and decreases to zero (see next paragraph) at some critical $R=R_c$. At that point and beyond, ionization can take place with zero energy change ($\Delta E=0$), which is a singular condition for Coulombic interaction amplitudes [16]. This singularity is key to the phenomenology of cloud chamber track formation.

To see directly why it's sensible that $\delta$ can vanish, consider that for large $R$ and $\varepsilon$, $g(R)$ in Equation (4.1) approaches $-Q^2/2R_i$. For $Q$=electron charge and $R_i$ = Bohr radius ~ 50pm, this is the hydrogen ground state binding energy, -13.6eV, of the same order of magnitude as, but deeper than, the ionization potentials of typical cloud chamber vapor molecules ethanol (10.5 eV [17]) and $H_2O$ (12.6eV [18]). Reference [19] tabulates Kelvin-Thomson radii for various ions in supersaturated n-butanol. The smallest radius tabulated ($K^+$) is of the same order of magnitude as the Bohr radius, which is encouraging. The mismatch is a factor of 3; but that doesn't undercut the hypothesis that $\delta$ can vanish in cloud chambers by $g(R)$ overtaking $E_I$, because a potassium atom is much bigger than the constituent atoms of ethanol or water.

Coulomb scattering from degenerate states

To begin understanding the singularity at $R=R_c$, let us examine the total cross section of this ionization interaction in more detail. It may seem strange here to turn to total cross section, since the concept is usually associated with interaction probability, while I have set myself the goal of *deriving* the Born spatial probability rule rather than simply reasserting it. As we shall see, I will eventually use total cross section as a device for reasoning about |emitted-particle wavefunction|$^2$ *without* presuming an a priori connection with probability. I can do this because, in the time-independent formulation of scattering theory [16], a scattering center for an inelastic transition modifies an initial plane wave $\alpha e^{i\mathbf{k}\cdot\mathbf{x}}$ by adding a scattered wave with far-field form $\alpha M e^{ikr}/r$, where $M$ is a solid-angle-dependent matrix that connects incoming to outgoing channels and $\alpha$ is an arbitrary complex multiplier. In this formulation, with or without a probability interpretation, the total rate at which square-norm escapes to infinity in the scattered wave is simply $\sigma s|\alpha|^2$, where $\sigma$ is the total cross section for this inelastic channel, calculated in the usual way. Typically, the initial plane wave is an approximation for the local behavior of a normalized (elastic) incoming wavefunction, in which case $\alpha$ is the local value of the wavefunction at the scattering center and, by total square-norm conservation, $\sigma s|\alpha|^2$ is the rate at which square norm drains out of the normalized elastic channel.

Following Ref. [20], the differential cross section for single-electron ionization by Coulombic interaction with a passing charged projectile is

$$d\sigma_{fi} = 8\pi Q^2 s^{-2} \left| F_{fi}(\mathbf{q}, E_e, d\Omega_e) \right|^2 q^{-3} dq dE_e d\Omega_e, \tag{4.3}$$

where $i$ and $f$ denote initial and final states; $E_e$ and $\Omega_e$ are outgoing electron kinetic energy and solid angle, respectively; $\hbar\mathbf{q}$ is the difference between outgoing and incoming projectile momentum; and, as before, $Q$ and $s$ are projectile charge and speed, respectively, assuming small energy loss $\Delta E$. This assumes the initial and final projectile states are plane waves. The initial plane wave is a stand-in for the local behavior of the persistent state wavefunction at the molecule to be ionized. The final plane wave is a stand-in for the local behavior of a free-state wavefunction at the molecule to be ionized. As long as the initial and final plane waves have different momenta, the final plane wave can only be a proxy for a free state, since there are many free states but only a unique persistent state.

In the Born approximation, the form factor $F_{fi}$ is given by a sum of overlap integrals

$$F_{fi} = \Sigma_l \langle \chi_f | e^{i\mathbf{q} \cdot \mathbf{r}_l} | \psi_i \rangle, \tag{4.4}$$

where $\psi_i$ is an initial state of droplet and no ion; $\chi_f$ is a final state of droplet and ion, with ejected electron characterized by $E_e$ and $\Omega_e$; and the sum is over the coordinates of all non-projectile constituents weighted by their electric charges. The singularity at $q=0$ is obvious in Equation (4.3). Naively, one might expect the singularity to be moderated by initial/final state orthogonality, which would make the overlap integrals in Equation (4.4) vanish for $\mathbf{q}=0$. But we know [21] this is generically *not* the case for ionization processes. What *does* moderate the singularity is the integral over $E_e$ and $\Omega_e$ when turning Equation (4.4) into a total cross section according to Equation (4.3). For $\delta=0$ and small $q$, the integral of $dE_e d\Omega_e$ goes to zero like $q$, so the singularity in cross section vs. $q$ alone is $q^{-2}$, and this singularity drives integrated cross section to scale like $q_{min}^{-1}$, where $q_{min}$ is the smallest $q$ possible. For small positive $\delta$ this amounts to scaling like $\delta^1$, which in turn is the same as scaling like $(R_c-R)^{-1}$.

Probability of first-droplet formation

The upshot of the last subsection is that, at a barely subcritical droplet, the ionization process drains total square norm from the persistent state into the ionized channel at the approximate rate $\sigma s|\psi|^2$, where $\psi$ is now the emitted-particle persistent-state wavefunction at the molecule to be ionized. [We have probably assumed implicitly that the persistent-state wavefunction and the cloud chamber medium are not entangled; this isn't unreasonable, since the decay source doesn't "know about" the detector as it leaks out the emitted-particle wavefunction.] The persistent state drains *entirely* into a free state combined with a liberated electron if $\sigma s|\psi|^2 \tau > 1$ ("suddenness" in Section 2), where $\tau$ is the (presumed short) time needed for the triggering subcritical droplet to evaporate (i.e. at time greater than $\tau$, there's no droplet left to do any draining). We can say this because unitary quantum mechanics conserves the square norm of the whole-system quantum state, and we can assume, as usual, that square norm is normalized to unity. Since, for $R$ close to $R_c$, the cross section takes the form $A/(R_c-R)$, the probability that the emitted particle drives the subcritical droplet "over the edge" by creating an ion is the probability that

$$As\tau|\psi|^2 > R_c - R. \tag{4.5}$$

Now recall that R is the radius to which the subcritical droplet would grow *if* there were no ionizing projectile to drive it critical. As a result, the probability in question – the probability of excess droplet formation due to an ionizing projectile – per unit volume and time, must be

$$\rho As\tau|\psi|^2, \tag{4.6}$$

i.e. the Born rule in position space ("probability distribution" in Section 2). This is for the first droplet. Applying Mott's reasoning [8], this ionization event transforms the persistent emitted particle wavefunction into a collimated free-particle wavefunction emanating from the point of nucleation, guaranteeing that subsequent droplets form a linear track. It is beyond the scope of this work to estimate numerical values for $\rho$, $A$ and $\tau$, but that must be done at some point to confirm that the foregoing actually makes experimental sense in all respects.

[For further work, an interesting puzzle: If there's a collection of, say, $U^{238}$ nuclei in a cloud chamber, why don't they all produce alpha decay tracks at the same time when "the right" subcritical droplet comes along?]

The foregoing rests on a tacit assumption that must now be made explicit. I wrote above about an amplitude of a particular state draining entirely into another state. Of course, this can never be literally true to infinitely many decimal places. What I'm really saying is that the amplitude of the initial state gets small enough that it no longer has any practical significance for observation. Essentially, I'm implicitly equating "small enough amplitude" with "doesn't matter," without quantifying "small enough." (I made the same assumption, more explicitly, in Reference [6].) It seems that somehow, to explain away quantum measurement axioms and account for experiments without rejecting other things we already know, we have to identify some sort of physical meaning – even if not probabilistic – for the size of a quantum amplitude. In this connection, I quote from Reference [6]:

"One must consider that an observer is himself made of wavefunctions. Under Hamiltonian evolution (Schroedinger's equation) wave functions can spread out into space or coalesce into bound dynamic entities. Those coalesced entities can be exceedingly complex, to the point of stumbling upon their own versions of thinking, acting and measuring, i.e. becoming us. The only fundamental metric available for distinguishing one such being from another, or from its surroundings, is wavefunction square norm. "Things" are concentrations of square norm. A big square norm should somehow translate into a coalesced entity's experience. It is hard to say how this works for square norms anywhere on a spectrum between 0 and 1, but for square norm very close to 0 or very close to 1 it should be simple: approximately 0 means nothing happens, approximately 1 means something always happens. [This picture is "philosophical" in quotes] because in principle you could run a huge computer simulation and see for yourself the simulated wavefunctions coalescing, then learning how to think and measure, and finally measuring."

So this assumption – that small wavefunction means small physical effect without an explicit wavefunction probability interpretation – seems to be a necessary condition for accounting for canonical quantum measurement behavior without explicit measurement axioms. I doubt that it could also be sufficient. The explicit reliance in this paper on Coulomb interactions undercuts this notion.

## 5. Detectors that rely on gas-filled enclosures and electrified conductors

Such detectors have a lot in common with cloud chambers. A particle (charged or otherwise) creates a track of ions in an enclosed gas, and a strong electric field draws liberated electrons to an anode that registers a voltage spike. As with cloud chambers, Mott's wave mechanics argument both explains why these ions create a linear track, and fails to explain the first ion as an event localized in space and time. But by contrast with cloud chambers, gas-and-voltage detectors don't have random near-critical structure formation, to be pushed critical by weak projectile wavefunctions.

However, what such detectors *do* have are loci characterized by a potential energy almost identical to Equation (4.1). Indeed, the almost identical expression $-Q^2/2R_i$ describes the image-charge potential induced by an ion of charge $Q$ at distance $R_i$ above the surface of a perfect conductor. For this reason, I conjecture that the first ion in a gas-and-voltage detector track forms close to an electrode, caused by a balance between ionization energy and image-charge potential, leading to a singular scattering condition, and then to an ion formation probability proportional to |projectile wavefunction|$^2$ at the site of ionization. At present, my own command of metallic surface physics is not informed enough to take this argument further.

## 6. Bubble chamber

Bubble chambers are not merely "negatives" of cloud chambers. Bubbles and droplets may share similar energetic competition between surface tension and volume energy. But the mechanisms that make passing charged particles form them are dramatically different. A droplet in a cloud chamber nucleates around a single ionized molecule in supercooled gas, while a bubble in a bubble chamber results from an "explosion" caused when a fast electron ejected from an ionized molecule deposits a large energy spike into a small volume of superheated liquid [22]. Moreover, the dielectric constant for a bubble chamber liquid is relatively close to unity, so any polarization-induced potential, as in Equation (4.1), would be insufficient to compensate for the energy required to ionize a liquid molecule. Yet we know [23] that particles emitted by slow decays do leave bubble chamber signatures.

So where could a passing charged particle induce a process that could have a singular cross section at zero momentum transfer? One place to look is the molecules on the surfaces of near-critical bubbles that form in a superheated liquid by routine thermal fluctuation. It takes nonzero energy transfer for a passing charged particle to eject such a molecule into the bubble interior, since it must overcome the attraction that binds the molecule to the bubble surface. But as a result, the surrounding liquid is free to pull in on itself slightly (that's how the bubble expands), possibly providing a compensating increase in bulk binding energy, leading to a singular degenerate scattering process as in Section 4.

## 7. Discussion

I have formulated a mechanism for how the Hamiltonian structure of quantum decay, the physics of droplets in supersaturated vapors, and the mathematics of quantum Coulomb scattering from degenerate states can together account for the observed phenomenology of track origination in cloud chambers, without having to invoke measurement axioms. I have speculated about how an analogue of the same mechanism could account for quantum measurement behavior in detectors that rely on gas-filled enclosures and electrified conductors such as ionization chambers, proportional counters and photomultiplier tubes. I have also speculated about track origination in bubble chambers.

I have shown how this phenomenology should conform to an emergent position-space Born rule, but there is actually no systematic experimental data on cloud chambers to confirm this conclusion. This is a glaring hole. To remedy this deficiency, someone should carefully tabulate how the starting points of decay tracks are distributed – in solid angle *and* distance – around a radioactive sample in a cloud chamber.

It might be possible to validate some of the theory in this paper by observing charged-particle tracks in cloud chambers made with nonstandard ingredients. For example (not a practical one, given the obvious hazards), if HF were the condensing vapor, then first droplets and therefore tracks might be dramatically suppressed because the ionization potential of HF (16eV [24]) would be too large to be overwhelmed by $g(R)$ and therefore would make it impossible for $\delta$ to vanish. Perhaps there are less dangerous alternatives. Following Section 5, it would be very interesting to learn how an experiment could show that slow-decay-generated ion tracks in ionization chambers or proportional counters or photomultiplier tubes always originate near electrodes.